\documentclass[prb,twocolumn]{revtex4}

\usepackage[latin1]{inputenc}
\usepackage[T1]{fontenc}
\usepackage[francais,english]{babel}
\usepackage{times}
\usepackage{graphicx}

\def\CoIII{Co\textsuperscript{3+}}
\def\CoIV{Co\textsuperscript{4+}}
\def\NaXXIII{{\textsuperscript{23}Na}}
\def\CoLIX{{\textsuperscript{59}Co}}
\def\NaxCoOO{Na$_x$CoO$_2$}
\def\NaCoOO{Na$_1$CoO$_2$}

\begin{document}
\title{Evidence of a single nonmagnetic Co\textsuperscript{3+} state in the Na$\mathbf{_1}$CoO$\mathbf{_2}$ cobaltate}
\author{G. Lang,\textsuperscript{1} J. Bobroff,\textsuperscript{1} H. Alloul,\textsuperscript{1} P. Mendels,\textsuperscript{1} N. Blanchard,\textsuperscript{1} and G. Collin\textsuperscript{2}}
\affiliation{
\textsuperscript{1}Laboratoire de Physique des Solides, UMR 8502, Université Paris-Sud, 91405 Orsay, France \\
\textsuperscript{2}Laboratoire Léon Brillouin, CE Saclay, CEA-CNRS, 91191 Gif-sur-Yvette, France
}

\begin{abstract}
Macroscopic magnetization, muon spin rotation ($\mu$SR), and NMR measurements were carried out to study magnetism in the Na$_1$CoO$_2$ cobaltate.
Using superconducting quantum interference device measurements, \NaCoOO\ is shown to have a bulk magnetic susceptibility much lower and flatter than that of \NaxCoOO\ with $x$=0.7--0.9.
In fact, $\mu$SR yields a signal of mostly nonmagnetic origin, which is attributed to the $x$=1 phase.
The intrinsic cobalt spin susceptibility corresponding to this $x$=1 phase is measured using \NaXXIII\ NMR. It is indeed found to be almost zero, in agreement with a low-spin 3+ charge state of all cobalt atoms.
This single state of Co ions in CoO$_2$ planes is confirmed by \CoLIX\ NMR, whose determination of cobalt shift and quadrupolar parameters allows us to give a reference value of the \CoIII\ orbital shift in cobaltates. \\

\end{abstract}

\maketitle

\section{Introduction}

Layered cobaltates have garnered much attention for their unusually high thermoelectric power and following the recent discovery of superconductivity in the Na$_{0.35}$CoO$_2$$\cdot$1.3H$_2$O phase.\cite{Takada}
In the case of the \NaxCoOO\ series, the various phases feature alternating layers of CoO$_6$ octahedra and sodium.
The edge-sharing CoO$_6$ octahedra result from the superposition of triangular lattices of oxygen and cobalt.
This opens up the possibility of geometric frustration of the interactions, such as between magnetic moments on cobalt sites.
Unlike cuprates, doping in \NaxCoOO\ cobaltates is inhomogeneous, with the most simple picture being that of a varying ratio of \CoIII\ and \CoIV\ depending on the sodium content.
The crystalline field forces the cobalt into a low-spin state, so that there are five (for \CoIV) or six (for \CoIII) electrons in the three lower $3d$ orbitals ($t_{2g}$ orbitals).
This results in respective spin $S$=1/2 and 0 for \CoIV\ and \CoIII.
Using as a reference the hypothetical sodium-empty situation ($x$=0) or the $x$=1 situation, the system can be seen as either an electron-doped Mott insulator or a hole-doped band insulator.

Depending on the sodium content and temperature, a variety of well-defined phases has been observed.
For $x$ in the 0.75--0.9 range, it has been shown\cite{Sugiyama075,Motohashi075,Bayrakci082,MendelsCascade,Sugiyama090PRB} that there are magnetic transitions with transition temperatures between 19 and 27~K, with the magnetic order encompassing up to the totality of the sample.
In these phases, even though the magnetic moments get increasingly diluted with higher $x$, magnetism remains present with an evolution of transition temperatures that is seemingly uncorrelated to the sodium content.
This raises the question of the actual charge and spin distribution.
Even in the simplest picture of an $x$-dependent ratio of \CoIII\ and \CoIV\ ions, one may expect a specific arrangement, such as triangular (\CoIV) and hexagonal (\CoIII) lattices\cite{MukhamedshinChargeOrder} for $x$=0.66, and a \CoIII\ kagome lattice\cite{BernhardPRL93} for $x$=0.75.
But there is also experimental evidence that points to intermediate charge states. For $x \approx$ 0.7, Mukhamedshin {\it et al.}\cite{MukhamedshinCoValences} present evidence of 3+, 3.3+, and 3.7+ cobalt charge states.

In the $x$=1 phase, from examining only the electronic structure, one would expect to have only \CoIII\ ions, without the possibility of charge disproportionation which would result in intermediate valences between 3+ and 4+.
There should then be no magnetism from the cobalt electrons, in contrast to the nearby $x$=0.75--0.9 phases.
In this case, \NaCoOO\ should be well suited to the determination of characteristic \CoIII\ parameters in \NaxCoOO.
These parameters can then be used as a reference point in the analysis of compounds of lower sodium content, to help in the attribution of charge and spin states to the different observed cobalts.
A previous magic-angle spinning NMR study by Siegel {\it et al.}\cite{Siegel} allowed determination of Na and Co NMR parameters at room temperature. But no temperature-dependent magnetic measurements were performed up to now in order to check for the actual magnetic and charge state of Co planes in \NaCoOO.

Here, we present a detailed study of the magnetic properties of Na$_{1}$CoO$_{2}$ using both macroscopic and local probes.
In Sec. \ref{sec:crystallo}, we give the crystallographic characteristics of the $x$=1 phase.
In Sec. \ref{sec:bulk}, using superconducting quantum interference device measurements, macroscopic susceptibility is shown to be essentially very small as compared to the $x$=0.7--0.9 phases, pointing toward no magnetism within the cobalt planes.
In Sec. \ref{sec:musr}, we present muon spin rotation ($\mu$SR) measurements done down to $T$=1.3~K which reveal the presence of a nonmagnetic component together with a smaller magnetic contribution.
NMR of both Na and Co nuclei is presented in Sec. \ref{sec:rmn}.
Na NMR allows us to show that the $x$=1 phase corresponds solely to the nonmagnetic contribution seen in both SQUID and $\mu$SR, while the magnetic contributions are due to spurious phases.
The cobalt plane susceptibility is then measured using the sodium shift and found to be almost zero, in agreement with a low-spin $S$=0 \CoIII\ state of each cobalt ion.
We are then able to measure the Co NMR signal corresponding to this $x$=1 phase, which shows accordingly a single cobalt site.
Its shift measures temperature-independent orbital contributions which show very good consistency with a previous measurement\cite{MukhamedshinCoValences} at $x \approx$ 0.7.

\section{Crystallographic structure}
\label{sec:crystallo}

\NaCoOO\ has been previously synthesized and its structure determined on both powder samples\cite{Fouassier} and single crystals.\cite{Takahashi}
We used \NaCoOO\ powder prepared by solid-state reaction of a stoichiometric mix of Co$_3$O$_4$ and Na$_2$O$_2$.
Synthesis took place at 700 $^{\circ}$C for 10~h under argon flux.
X-ray powder diffraction data were recorded in the range 2$\theta$=10$^{\circ}$--130$^{\circ}$, using the Cu $K\alpha$ line.
A rhombohedral structure of space group $R\bar{3}m$ (\#166) is found, with three formula units per unit cell.
The atomic positions in the unit cell are given in Table~\ref{tbl:crystallo}.
Rietveld refinements gave lattice constants $a$=2.888 26(5)~\AA\ and $c$=15.5989(2)~\AA, with $R$=4.19 and $wR$=6.93.
No additional satellite reflections were observed.
The absence of any modulation or distortion is indicative of a perfect long-range ordering at the time of synthesis.

\begin{table}[htbp]
\begin{tabular}{ccccccc}
\hline
\hline
Atom & Site & Occupancy & {\it x} & {\it y} & {\it z} & {\it B} \\
\hline
Na & 3{\it a} & 1.019(9) & 0 & 0 & 0 & 0.85(8) \\
Co & 3{\it b} & 1 & 0 & 0 & 0.5 & 0.28(3) \\
O & 6{\it c} & 1 & 0 & 0 & 0.7697(2) & 0.58(8) \\
\hline
\hline
\end{tabular}
\caption{Structural atomic parameters in \NaCoOO\ at $T$=300~K.}
\label{tbl:crystallo}
\end{table}

\begin{figure}[htbp]
\includegraphics[width=7cm]{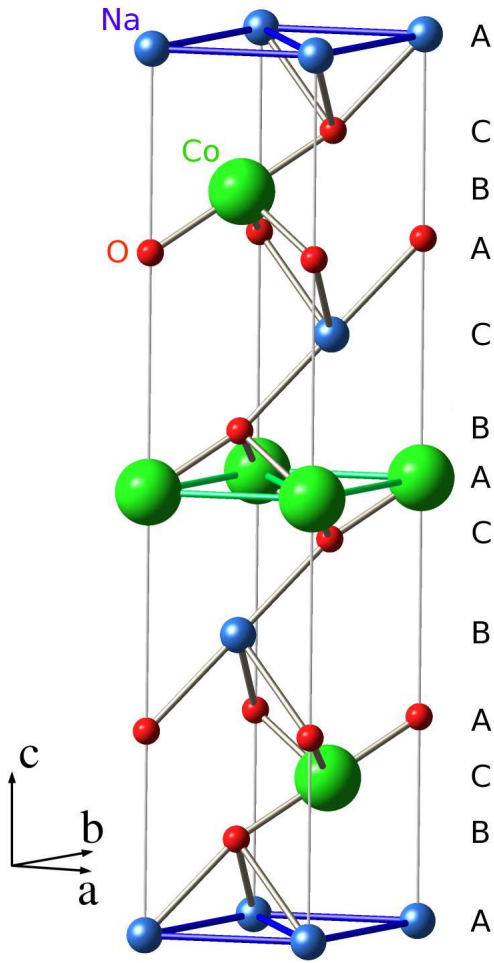}
\caption{(Color online) Unit cell of the \NaCoOO\ phase. Selected bonds are drawn for clarity of the structure, without consideration of physical relevance.}
\label{fig:struct}
\end{figure}

As shown in Fig.~\ref{fig:struct}, the successive atomic layers along the $c$ axis follow an ...-A-B-C-... close-packing-like configuration.
A peculiarity of the \NaCoOO\ structure is that there is only one crystallographic site for sodium atoms, which are all at the barycenters of the higher and lower cobalt triangles (Na2 site).
No sodium is to be found directly above and below cobalts, which is the other usual sodium site (Na1 site) in \NaxCoOO\ cobaltates.
This contrasts with other \NaxCoOO\ phases,\cite{HuangCoupling} such as for $x$=0.5 where there is equal occupancy of Na1 and Na2 sites.\cite{Huang05}

As a consequence of this single sodium site, there is also only one crystallographic site for cobalt.
In the above picture of a single \CoIII\ charge state of all cobalt atoms, one may then expect a single environment for all sodium nuclei, and for all cobalt nuclei.

\section{Bulk magnetic susceptibility}
\label{sec:bulk}

\begin{figure}[htbp]
\includegraphics[width=6.5cm]{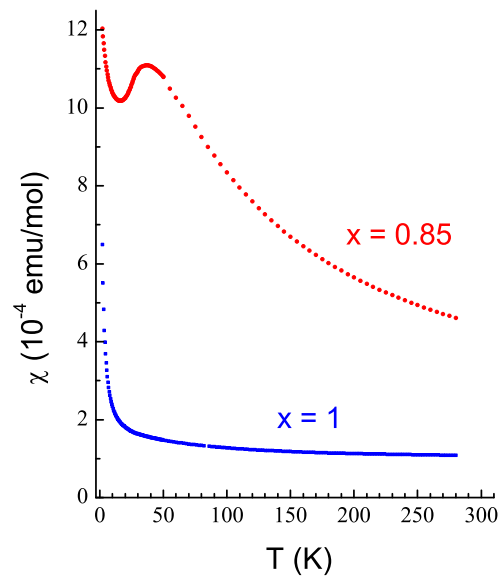}
\caption{(Color online) Temperature dependence of the bulk magnetic susceptibility of \NaCoOO\ as measured by SQUID in $H$=1000~G, compared to that of the $x$=0.85 phase.
In the $x$=0.85 phase, the cusp and decrease below $T \approx$30~K reveal the presence of long-range magnetic ordering seen in $\mu$SR.\cite{MendelsCascade}
The SQUID data at $x$=0.85 were obtained on the sample used in that $\mu$SR study.}
\label{fig:squid}
\end{figure}

Although \NaCoOO\ lies next to the magnetism-rich $x$=0.75--0.9 zone in the \NaxCoOO\ phase diagram, we expect from the electronic structure that no magnetic cobalt ions are present.
Macroscopic magnetization measurements allow us to probe the bulk magnetic susceptibility, which is essentially the total cobalt electronic susceptibility.
Magnetization was measured using a SQUID Quantum Design MPMS Magnetometer.
We found no hysteresis, to be caused by ferromagnetism, in the field dependence of magnetization.
Bulk magnetic susceptibility was then measured in a 1000~G applied field in the 2--300~K range as shown on Fig.~\ref{fig:squid}.
\NaCoOO\ exhibits a small and flat susceptibility, with a significant increase only at low temperature.
This low-temperature tail has been observed at all other compositions as illustrated on Fig.~\ref{fig:squid} for $x$=0.85, and is usually attributed to spurious paramagnetic local moments.
Here the low-temperature increase is fitted to a Curie law, with a Curie constant which would correspond to 0.3\% of $S$=1/2 paramagnetic impurity per cobalt site.
We will demonstrate hereafter, using sodium NMR, that this increase is due to impurities extrinsic to the $x$=1 phase.
The flat behavior at higher temperature is very different from that of the $x$=0.85 data shown on Fig. \ref{fig:squid}.
Indeed, in the $x$=0.7--0.9 range, the susceptibility displays a $1/(T+\Theta)$ Curie-Weiss behavior with values at least four times larger than our present measurement.
The comparatively small value of the $x$=1 susceptibility then points toward the absence of large Curie-Weiss paramagnetism in the cobalt planes, as will be discussed in Sec. \ref{sec:rmn}.
The susceptibility would then be the sum of the diamagnetic $\chi_{dia}$ and orbital  $\chi_{orb}$ contributions.
$\chi_{dia}$ is estimated\cite{MukhamedshinCoValences} in the $x \approx$ 0.7 phase at -0.35$\times$10$^{-4}$~emu/mol, so that our data then yield $\chi_{orb} \approx$ 1.5$\times$10$^{-4}$~emu/mol, to be compared in Sec. \ref{sec:rmn} to the cobalt orbital shift.
The phase transition to static long-range ordered magnetism shown by some of these $x$=0.7--0.9 phases at $T$$<$30~K appears as a kink on SQUID measurements, as in the usual antiferromagnets.
For $x$=1, we observe two similar but much smaller kinks (not directly visible on Fig. \ref{fig:squid}) around $T$=20 and 27~K, the exact same transition temperatures previously observed in the $x$=0.82 and 0.85 compositions.\cite{Bayrakci082,MendelsCascade}
These two phases are estimated to make up 3--5\% of the sample mass.

\section{$\mu$SR measurements}
\label{sec:musr}

In order to check that the $x$=1 composition shows no intrinsic magnetic order, a more local probe is required.
Therefore we use muon spin rotation, which is highly sensitive to local static magnetism as demonstrated in previous studies.\cite{Bayrakci082,MendelsCascade,Sugiyama090PRB,Sugiyama090PRL}
Fully spin-polarized positive muons $\mu^+$ are implanted in the first few hundred micrometers of the sample, where they stop at specific sites, likely forming an $\approx$1~\AA\ $\mbox{(O--$\mu$)}^-$ bond.
As the muon has a $S$=1/2 spin, it is sensitive to the local magnetic field $H_{loc}$ with its spin precessing at the Larmor frequency $\omega_{\mu} = \gamma_{\mu} H_{loc}$, where $\gamma_{\mu}$ is the gyromagnetic ratio of the muon (2$\pi \times $13.6~kHz/G).
Having a half-life of 2.2~$\mu$s, most of the muons decay in the first dozen microseconds after implantation, with each of them emitting a positron preferentially in the direction of the muon spin at the time of decay.
Since the implanted muons are initially fully spin polarized, this means angularly resolved positron detectors can measure the time evolution of the total muon spin polarization, or signal asymmetry $\cal A$, in the sample.
In the case of a two-detector setup (forward + backward) it is written as:
$${\cal A}(t) = \frac{N_B(t) - \alpha N_F(t)}{N_B(t) + \alpha N_F(t)}$$
where $N_B$ and $N_F$ are the instantaneous number of events at each detector at time $t$ and $\alpha$ is a setup-dependent correction factor.
From $\cal A$ one can deduce information about $H_{loc}$.
In a zero-field (no applied field) measurement, significant static magnetism should lead to a fast decrease of $\cal A$, or to a decaying set of oscillations if there is magnetic ordering.

\begin{figure}[htbp]
\begin{center}
\includegraphics[width=7cm]{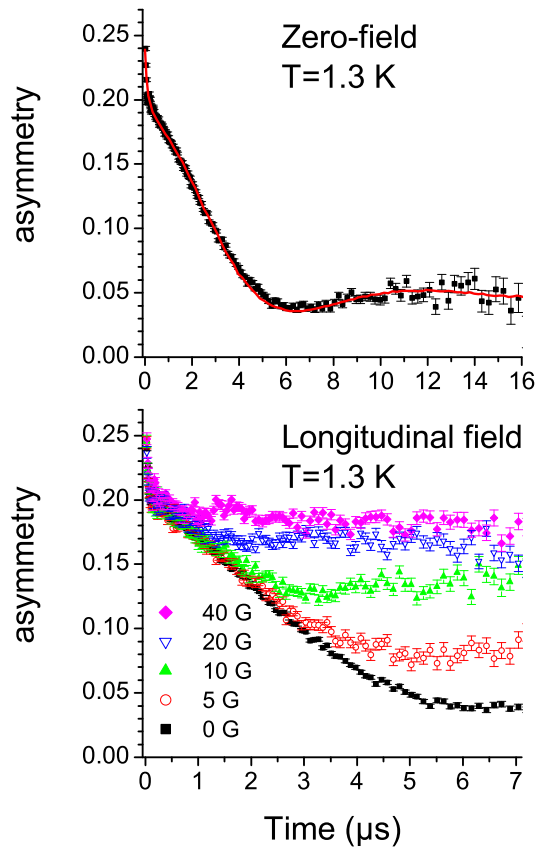}
\end{center}
\caption{(Color online) $\mu$SR signal at $T$=1.3~K. In the upper panel, zero field signal with a three component (dynamic Kubo-Toyabe, stretched exponential, and a background) fit shown as a full line.
In the lower panel, similar experiment with various applied longitudinal fields, showing the decoupling from the nuclear dipolar field.
}
\label{fig:musr}
\end{figure}

The measurements were carried out on the MuSR experimental setup at the ISIS facility.
Samples were mounted on a copper sample holder, with a silver back plate.
The upper panel of Fig.~\ref{fig:musr} shows the asymmetry versus time in zero field, at $T$=1.3~K which is significantly smaller than the lowest observed transition temperature ($\approx 19$~K) in the $x$=0.75--0.9 region.
We observe two contributions to the signal.
The largest one (about 70\%) features a slow relaxation on the order of 5--6~$\mu$s with a Gaussian-like behavior, indicating that the corresponding muons see no static magnetism.
The only relaxation source is then the nuclear dipolar field.
To ensure the dominantly static character of this field, we proceeded to longitudinal field measurements.
By applying a field $H_{LF}$ directed along the initial muon polarization, it is indeed possible to check whether the local field $H_{loc}$ is essentially a small static field due to the nuclear dipoles.
The lower panel of Fig.~\ref{fig:musr} shows the effect of various $H_{LF}$ values on the asymmetry at $T$=1.3~K.
Fields close to 40~G are enough to maintain the muon polarization along its $t$=0 initial direction, for the muons implanted in the nonmagnetic part of the sample.
This confirms that this contribution reflects the effect of a quasistatic nuclear dipolar field distribution with an amplitude $\approx$ 5--10~G, similarly to the phases of lower $x$ in the paramagnetic state.
It is then consistent to describe the zero field relaxation by a Kubo-Toyabe relaxation function,\cite{KuboToyabe} possibly with some limited time fluctuations of the angular distribution of nuclear dipoles (``dynamic'' Kubo-Toyabe).
Besides, the nonmagnetic fraction of the sample can also be determined by applying a small transverse field $H_{TF}$ (i.e., perpendicularly to the initial muon polarization) and measuring the resulting oscillating component of the asymmetry. Using $H_{TF}$=20~G, we find the same nonmagnetic volume fraction as in zero field measurements, with no significant variation in the $T$=1.3--90~K range.
The second contribution to the signal (about 30\%) undergoes fast relaxation during the first microsecond.
This sharp spin depolarization is due to magnetism of electronic origin, in contrast to the first contribution.
Measurements at higher longitudinal fields than in Fig. \ref{fig:musr} showed the predominantly static character of this magnetic contribution (data not shown).
Most of it is still present up to $T$=90~K, consistently with the aforementioned weak transverse field measurements.
Therefore this contribution cannot be attributed to any of the $x$=0.75--0.9 phases, whose magnetic ordering develops in the $T$=19--27~K range.
It is then more likely to be due to some spurious cobalt oxide. For example, CoO is an antiferromagnet with $T_N$=291~K.
No clear signature is obtained on the x-ray diffraction pattern, so that in the case of a stoichiometric oxide such as CoO the long-range order has to be poor.
Overall, this contribution might suggest a small degradation of our sample after synthesis.
In light of the very low magnetic susceptibility as measured by SQUID, we find it hard to determine the actual nature of this spurious magnetic phase.
For instance, literature values for CoO susceptibility\cite{UchidaCoO} show that 30\% of CoO in volume (as $\mu$SR yields volume fractions) would contribute a susceptibility per sample unit mass more than twice higher than the total observed susceptibility.

In order to further characterize the relaxation, we fitted the data with a sum of a dynamic Kubo-Toyabe relaxation function $\Phi_{DKT}$ corresponding to the nonmagnetic part of the asymmetry, a stretched exponential corresponding to the static magnetic part, and a background $A_b$:
\[
{\cal A}_{fit}(t) = A_{DKT} \Phi_{DKT}(\Delta H, \nu, t) + A_e e^{-(\frac{t}{\tau})^\alpha} + A_b
\]
The resulting fit is shown in the upper panel of Fig.~\ref{fig:musr} as a full line.
The dynamic Kubo-Toyabe function has an asymmetry $A_{DKT}$=0.154$\pm$0.006, with a width $\Delta H$=3.50$\pm$0.15~G of the Gaussian local field amplitude distribution, and a small nuclear dipolar field fluctuation frequency $\nu$=0.09$\pm$0.02~MHz.
The stretched exponential has an asymmetry $A_e$=0.08$\pm$0.01, a time constant $\tau$=0.095$\pm$0.020~$\mu$s, and an exponent $\alpha$=0.50$\pm$0.05, while $A_b$=0.025$\pm$0.003.
$\Delta H$ translates to a typical $\sigma \approx$0.30~$\mu$s\textsuperscript{-1} Gaussian width in the frequency domain.
Simulation indicates that these $\Delta H$ and $\nu$ values are in good agreement with the above longitudinal field measurements.

\section{NMR measurements}
\label{sec:rmn}

One can wonder whether the low-temperature Curie tail observed in SQUID is intrinsic to the $x$=1 phase or caused by the presence of spurious phases.
To answer this question and probe the charge state of cobalt atoms in the nonmagnetic phase seen in $\mu$SR, we performed \NaXXIII\ NMR which measures the intrinsic cobalt planes susceptibility as shown in previous studies,\cite{MukhamedshinChargeOrder,MukhamedshinCoValences} as well as \CoLIX\ NMR which gives direct information on the cobalt charge state.
Indeed, NMR is a local probe that can distinguish unambiguously between different cobalt phases, as the explored sites are known without ambiguity, which is not the case with $\mu$SR.

\subsection{General NMR background}

When placed in an external magnetic field $\mathbf{H_0}$, an atomic nucleus of spin $\mathbf{I}$ has its spin energy levels determined by the Zeeman Hamiltonian ${\cal H}_Z = - \gamma \hbar \: \mathbf{I} \cdot \mathbf{H_0}$, where $\gamma$ is the nuclear gyromagnetic ratio.
The corresponding energies are given by $E = - m_I \gamma \hbar H_0$ with $m_I \in [-I, -I+1, \cdots, I-1, I]$, so that for $I \ne 0$, there are 2$I$+1 levels evenly spaced by $\gamma \hbar H_0$.
By applying radiofrequency at the Larmor pulsation $\omega_0 = \gamma H_0$, transitions can be triggered between consecutive levels, thus altering the average spin magnetization.
As it precesses around $\mathbf{H_0}$, it induces a signal in a receiver coil.

In a solid, the resonance frequency can be shifted as compared to its atomic value $\nu_0$, following:
\[
\nu = \nu_0 (1 + K_{orb} + K_{cs} + K_{spin})
\]
The orbital shift $K_{orb}$ reflects the orbital effect of valence electrons. It is usually temperature independent.
Similarly, the chemical shift $K_{cs}$ reflects the orbital effect of inner-shell electrons.
The spin shift $K_{spin}$ is proportional to the electronic spin susceptibility $\chi_{spin}$, due to hyperfine couplings between the nucleus and surrounding electrons, through
\[
K_{spin} = \frac{A_{hf}^{spin} \; \chi_{spin}}{{\cal N}_A \; \mu_B}
\]
where $\chi_{spin}$ is in emu/mol, $A_{hf}^{spin}$ is the spin hyperfine coupling constant, ${\cal N}_A$ is the Avogadro number, and $\mu_B$ is the Bohr magneton. 
As some of the contributions to $K$ are anisotropic, it is actually a tensor of eigenvalues $K_x$, $K_y$, $K_z$.

The two nuclei that are considered here, \NaXXIII\ and \CoLIX, have a nuclear spin higher than 1/2, with, respectively, $I$=3/2 and 7/2. This reflects the anisotropy of the nuclear charge, and results in sensitivity to the local electric field gradient (EFG), in addition to the orbital and spin susceptibility.
The total Hamiltonian becomes:\cite{Slichter}
\[
{\cal H} = {\cal H}_Z + \frac{e Q}{4I(2I-1)} \left[V_{zz}(3I_z^2 - I^2) + (V_{xx}-V_{yy})(I_x^2-I_y^2) \right]
\]
where $Q$ is the nuclear quadrupole moment and $V_{xx}/V_{yy}/V_{zz}$ are the eigenvalues of the EFG tensor.\cite{footnoteEFG}
From these, we define the three quadrupolar frequencies:
\[
\nu_i = V_{ii} \frac{3 e^2 Q}{2I(2I-1) h} \;\;\;\;\;\;\; (i = x,y,z)
\]
We then define $\nu_Q = \left| \nu_z \right|$ and the quadrupolar asymmetry $\eta = \left| \frac{\nu_y - \nu_x}{\nu_z} \right|$. $\nu_Q$ and $\eta$ fully determine the quadrupolar tensor.
In the high-magnetic-field limit ($\omega_0 \gg 2\pi \nu_Q$), treatment as a perturbation to ${\cal H}_Z$ applies.
The first-order contribution suppresses the degeneracy of the Zeeman transitions so that, along each principal direction of the EFG tensor, a single NMR resonance actually splits into 2$I$ resonances, evenly spaced by the corresponding eigenfrequency $\nu_x$, $\nu_y$, or $\nu_z$.
The spectrum is then further modified by the second-order contribution, which breaks slightly the even spacing between these resonances.

\subsection{Sodium NMR}

\begin{figure}[htbp]
\begin{center}
\includegraphics[width=7.3cm]{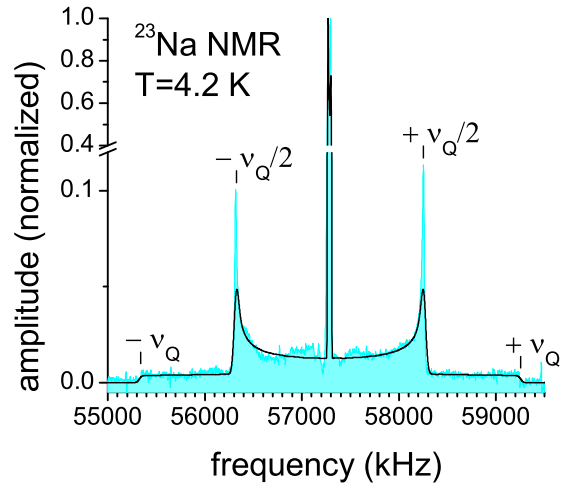}
\end{center}
\caption{(Color online) \NaXXIII\ NMR spectrum at $T$=4.2~K, with a fit of the powder distribution shown as a full line. The reference is at $\nu_0$=57.3~MHz. Ticks mark the four quadrupolar singularities of the powder distribution.}
\label{fig:Na_spectrum}
\end{figure}

Measurements are done in a $H_0$=0--7~T sweepable field as well as in a fixed $H_0$=7.5~T field, using a superheterodyne home-made spectrometer.
NMR echoes are obtained using $\frac{\pi}{2}$-$\tau$-$\frac{\pi}{2}$ pulse sequences; then Fourier transformation is applied.
We use $\frac{\pi}{2}$=3~$\mu$s and $\tau$=150~$\mu$s, with a typical recovery time between two consecutive sequences of 250~ms.
Note that the $T_1$ relaxation time does not increase by more than an order of magnitude from 85 down to 4~K, in spite of the lack of magnetic moments as demonstrated below.
We link this to the likely presence of defects in the sodium layers, due to the ease with which sodium is deintercalated at $x$=1.
Individual Fourier transforms of the signal are recorded at different field values; then the full spectrum is obtained by assembling together the Fourier transform spectra following the Clark {\it et al.} method.\cite{Clark}
All shifts are measured relative to a NaCl reference.
We show in Fig.~\ref{fig:Na_spectrum} the Na spectrum obtained at $T$=4.2~K.
We observe a narrow main transition on top of a broad distribution with four equally spaced singularities, symmetric with respect to the central line.
This is characteristic of a three-dimensional powder distribution.
Indeed, we use powder samples, made up of randomly oriented crystallites, so that the magnetic shift and electric quadrupolar tensors are randomly oriented with respect to the applied field.
This results in a continuous distribution of each resonance frequency.
The spectrum corresponds to a unique sodium site, as expected from the crystallographic structure.
The fact that the singularities are equally spaced indicates an axially symmetric EFG tensor ($\eta=0$).
This is consistent with the Na site having a threefold symmetry axis parallel to the $c$ axis, due to a unique cobalt charge state.
In that case, the quadrupolar frequency is equal to twice the peak spacing, yielding $\nu_Q \approx$1.96~MHz.
In addition, the symmetry of the singularities with respect to the central line shows that there is no significant magnetic shift anisotropy.
This is confirmed by the fact that the central lineshape is well accounted for by a purely quadrupolar distribution.
The full spectrum is well fitted by a powder distribution with $\eta$=0$\pm$0.02, $\nu_Q$=1.96$\pm0$.02~MHz, and an essentially isotropic magnetic shift ${^{23}K_{iso}}$=44~ppm, as shown by the full line on Fig.~\ref{fig:Na_spectrum}.
For higher temperatures up to $T$=300~K, the spectrum is found almost unchanged.
There is a less than 3\% increase of the quadrupolar frequency $\nu_Q$.
\begin{figure}[htbp]
\begin{center}
\includegraphics[width=6.5cm]{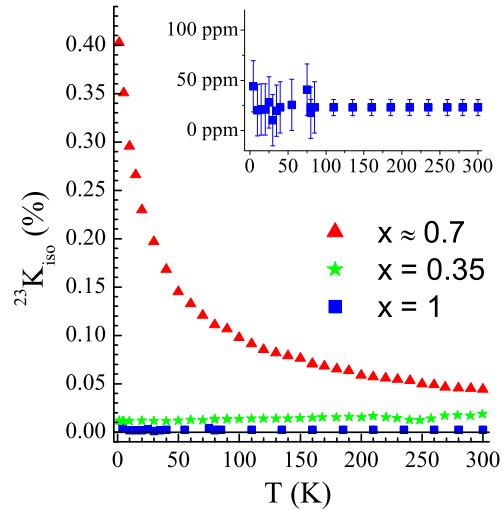}
\end{center}
\caption{(Color online) Temperature dependence of the isotropic magnetic shift of sodium in the $x$=1 phase, compared to the $x$=0.35 and $x \approx$0.7 phases (data taken from Ref. \cite{MukhamedshinChargeOrder}). Some of the same data are shown in more detail in the inset.}
\label{fig:Na_shift}
\end{figure}
Figure~\ref{fig:Na_shift} shows the temperature dependence of the isotropic part of the shift ${^{23}K_{iso}}$ for $x$=1, compared to the $x$=0.35 and $x \approx$ 0.7 phases.
The ${^{23}K_{iso}}$ value for $x$=1 is extremely low, with no measurable temperature dependence.
Note that the larger error bars below $T$=85~K are due to the error on the field value of the field-sweep setup.
In \NaxCoOO, sodium nuclei are coupled by large hyperfine couplings to neighboring cobalts through oxygen orbitals.\cite{MukhamedshinChargeOrder,Carretta}
These hyperfine couplings turn sodium nuclei into sensitive probes of the electronic spin susceptibility ${^{59}\chi_{spin}}$ of the nearby Co layers through
\[
{^{23}K} = \frac{{^{23}A_{hf}^{spin}} \; {^{59}\chi_{spin}}}{{\cal N}_A \; \mu_B} + {^{23}K_{cs}}
\]
where ${^{23}A_{hf}^{spin}}$ is the hyperfine coupling constant between one sodium and the two cobalt planes above and below.
This relation is well demonstrated at $x \approx$ 0.7, where the same temperature dependence for $\chi$ and ${^{23}K}$ is observed.\cite{MukhamedshinChargeOrder}
At such a composition, ${^{23}K_{cs}}$ is found to be much smaller than the first term, ${^{23}K_{spin}}$.
For $x$=1, our data show that ${^{23}K}$ is unusually low, of the order of ${^{23}K_{cs}}$ at $x$$<$1, so that ${^{23}K_{spin}}$ itself is very small.
Therefore the cobalt spin susceptibility ${^{59}\chi_{spin}}(T)$ is close to zero down to $T$=4.2~K.
This is strong evidence of a low-spin 3+ charge state of all cobalt atoms, for which no net electronic magnetic moment is expected.
Besides, if the low-temperature Curie tail observed in SQUID were intrinsic to the $x$=1 susceptibility, it should show up as well in the temperature dependence of ${^{23}K_{iso}}$, which it does not.
In contrast, if it corresponded to local moment impurities diluted in the $x$=1 phase, it should lead to at least a NMR dipolar broadening of the Na line. Using the hypothesis of 0.3\% of $S$=1/2 paramagnetic impurity per cobalt site calculated for this Curie tail in Sec. \ref{sec:bulk}, one would expect\cite{Walstedt} this broadening to follow $\Delta \nu = \frac{31}{T}$~kHz (with $T$ the temperature) if all these spins were in the $x$=1 phase.
This broadening is not observed.
The lack of temperature dependence of the linewidth in our measurements thus proves that the Curie tail corresponds to local moments extrinsic to the $x$=1 phase, i.e., in spurious phases.

\subsection{Cobalt NMR}

\begin{figure}[htbp]
\begin{center}
\includegraphics[width=7.5cm]{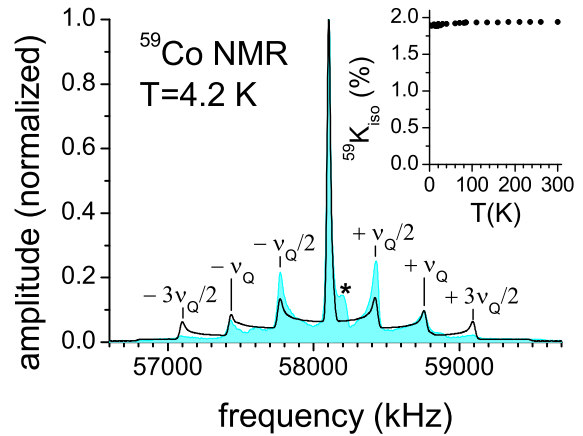}
\end{center}
\caption{(Color online) \CoLIX\ NMR spectrum at $T$=4.2~K, with a fit of the powder distribution shown as a full line. The reference is at $\nu_0$=57.02~MHz. Ticks mark the detectable quadrupolar singularities of the powder distribution. A star marks the central line of a spurious cobalt signal. The inset shows the temperature dependence of the isotropic part of the magnetic shift.}
\label{fig:Co_spectrum}
\end{figure}

Having established the nonmagnetic character of the $x$=1 phase, we now turn to \CoLIX\ NMR to characterize the charge state of cobalt atoms.
The experimental details are similar to \NaXXIII\ NMR except that the typical time between pulses is 50~$\mu$s, since some cobalts in \NaxCoOO\ have a short $T_2$ relaxation time.
The sequence repetition time is 100~ms or less, as the $T_1$ relaxation time again does not significantly increase at low temperature.
The \CoLIX\ NMR reference used is 10.054~MHz/T.
Figure~\ref{fig:Co_spectrum} shows a field-sweep measurement of the cobalt spectrum at $T$=4.2~K.
As in \NaXXIII\ NMR, the spectrum consists of a narrow central line on top of a broad distribution with singularities. The number of singularities is higher since the \CoLIX\ nuclear spin is $I$=7/2 instead of $I$=3/2 for sodium. Only six of the 12 expected powder singularities ($-3 \nu_Q$, $-\frac{5}{2}\nu_Q$, \dots, $+\frac{5}{2}\nu_Q$, $+3 \nu_Q$) are detected for sensitivity reasons and are shown in Fig.~\ref{fig:Co_spectrum}.
Similarly to \NaXXIII\ NMR, the quadrupolar singularities are equally-spaced and symmetric with respect to the central line.
The quadrupolar tensor is thus again axially symmetric ($\eta$=0) and the magnetic shift is mostly isotropic.
$\nu_Q$ is found equal to 0.67$\pm$0.02~MHz.
Some small magnetic shift anisotropy needs to be introduced to account for the central line width and shape.
This distribution pattern corresponds to a single cobalt site, in contrast to other compositions.
For instance, there are three different sites at $x \approx$0.7, resulting in three distribution patterns that yield a complex spectrum.
Here, the full spectrum is indeed well fitted using a single-site powder distribution with the above quadrupolar parameters and $K_x$=$K_y$=(1.88$\pm$0.02)\%, $K_z$=(1.96$\pm$0.03)\%.
The fit plotted as a full line on Fig.~\ref{fig:Co_spectrum} accounts for all singularities but not for the small signal marked with a star.
Note that this signal is artificially enhanced in this figure due to its shorter $T_1$ and longer $T_2$ relaxation times.
Measurements taking this into account show that it corresponds to 5--10\% of all cobalts.
It probably corresponds to another nonmagnetically ordered cobalt phase, possibly linked to the spurious phases observed in the SQUID measurements.
The main signal is characteristic of a single crystallographic cobalt site, with a single charge state.
It is safely associated with \CoIII-only sites in the $x$=1 phase.
Similarly to sodium, the spectrum is found almost temperature independent up to $T$=300~K, with a slight increase of the quadrupolar frequency to $\nu_Q$=0.69$\pm$0.02~MHz.
Likewise, at low temperature, there is no broadening of the central line that could be linked to the paramagnetic impurities causing the Curie tail, thus confirming their extrinsic character.
The temperature dependence of the isotropic shift ${^{59}K_{iso}}$ is plotted in the inset of Fig.~\ref{fig:Co_spectrum}.
It shows no measurable variation between $T$=1.2 and 300~K.
In a previous study\cite{MukhamedshinCoValences} at $x \approx$0.7, the ratio between ${^{59}K_{spin}^{iso}}$ and ${^{23}K}$ was found to be at most 2.6 for \CoIII\ sites.
As sodium NMR has shown that ${^{23}K} \approx$23~ppm, we then expect ${^{59}K_{spin}^{iso}}$ to be of the order of 60~ppm.
The measured cobalt shift is over two orders of magnitude larger than this and is then equal to the \CoIII\ orbital shift, which is temperature independent as expected.
Since the single \CoIII\ charge state results in a $c$-parallel threefold symmetry axis at the cobalt site, the $z$ axis of the magnetic shift and EFG tensors coincides\cite{footnoteTensors} with the $c$ axis, so that we can write $K_z$=$K_c$.
The orbital shift values ${^{59}K_{orb}^{iso}}$=1.91\% and ${^{59}K_{orb}^{c}}$=1.96\% are in remarkable agreement with the previous reference for $x \approx$0.7, where the 3+ charge state was associated with ${^{59}K_{orb}^{iso}}$=1.89\% and ${^{59}K_{orb}^c}$=1.95\%.
Therefore these values constitute a reliable reference for \CoIII\ ions in cobaltates.
Finally, by combining the upper bound to ${^{59}\chi_{orb}}$ given by SQUID measurements and the measured ${^{59}K_{orb}}$, a lower bound of $\approx$700~kOe can be extracted for the cobalt orbital hyperfine coupling constant, which is defined similarly to $A_{hf}^{spin}$.
The calculated value for the free \CoIII\ ion is 838~kOe,\cite{Freeman} so that this experimental lower bound is of reasonable magnitude, showing the rough quantitative consistency between SQUID and NMR measurements.

\section{Conclusion}

Using SQUID measurements, we have evidenced that the bulk magnetic susceptibility of our material is much lower and flatter than in other \NaxCoOO\ phases with $x$=0.7--0.9, except for a small low-temperature Curie tail.
Muon spin rotation showed that there are two contributions to the muon signal.
The largest one is that of a nonmagnetic phase and is due to \NaCoOO, while we link the other one to spurious magnetic phases.
By performing \NaXXIII\ NMR, it was confirmed that the low-temperature Curie tail is due to paramagnetic impurities extrinsic to the $x$=1 phase.
Note as well that the Co frozen moments observed in $\mu$SR do not coexist on the atomic scale with the nonmagnetic $x$=1 cobalt sites.
Using the presence of hyperfine couplings between sodium nuclei and the cobalt planes, it was shown that the cobalt spin susceptibility is very close to zero, which is consistent with an $S$=0 3+ charge state of all cobalt atoms.
\CoLIX\ NMR confirmed this single cobalt state and allowed determination of the magnetic shift value and anisotropy of the \CoIII\ ion, from which the orbital contribution was extracted.
In the study of other cobaltates, its value may prove useful in the identification of cobalt sites, since the charge state of cobalt atoms in such phases is usually not as clear as it is in \NaCoOO.
In conclusion, we clearly established the nonmagnetic low-spin \CoIII\ character of cobalt planes in the $x$=1 phase.
We also showed that local probes such as NMR and $\mu$SR are clearly needed to distinguish between intrinsic and spurious contributions in these compounds.

\section*{ACKNOWLEDGMENTS}

We thank A. D. Hillier, F. Bert, and A. Olariu for valuable help with the $\mu$SR experiments. This work was supported at ISIS by the EU Framework Programme VI.


\begin{references}
\bibitem{footnoteEFG} It is assumed in this expression that the $z$ axis is the symmetry axis of the nuclear charge. Note that the axes of the quadrupolar tensor may or may not coincide with the cell axes, which is also true of the magnetic shift tensor.
\bibitem{footnoteTensors} These two tensors possess the symmetries of the system giving rise to them, and the only way that they can include a threefold symmetry axis is by having such an axis along their main axis, together with a zero anisotropy.
\end{references}
\end{document}